\documentclass[12pt]{article}
\usepackage{amscd,amssymb,amsmath,latexsym,enumerate}
\usepackage[mathscr]{euscript}
\usepackage{epsfig}
\usepackage{fancybox}
\usepackage{verbatim}

\title{$\ZM_2$-indices and factorization properties \\ of odd symmetric Fredholm operators}

\author{Hermann Schulz-Baldes
\\
\\
{\small Department Mathematik, Universit\"at Erlangen-N\"urnberg, Germany}
}

\date{ }

\newtheorem{theo}{Theorem}

\newtheorem{proposi}{Proposition}

\newcommand{\BM}{{\mathbb B}}
\newcommand{\CM}{{\mathbb C}}
\newcommand{\NM}{{\mathbb N}}
\newcommand{\RM}{{\mathbb R}}
\newcommand{\SM}{{\mathbb S}}

\newcommand{\ZM}{{\mathbb Z}}
\newcommand{\PM}{{\mathbb P}}

\newcommand{\KM}{{\mathbb K}}

\newcommand{\UM}{{\mathbb U}}

\newcommand{\FM}{{\mathbb F}}

\newcommand{\Bb}{{\cal B}}

\newcommand{\Ff}{{\cal F}}

\newcommand{\Vv}{{\cal V}}

\newcommand{\Ee}{{\cal E}}

\newcommand{\Cc}{{\cal C}}

\newcommand{\Hh}{{\cal H}}

\newcommand{\one}{{\bf 1}}

\newcommand{\Ind}{{\rm Ind}} 
\newcommand{\Ran}{{\rm Ran}} 
\newcommand{\Ker}{{\rm Ker}}

\newcommand{\ess}{{\mbox{\rm\tiny ess}}}

\begin{document}

\maketitle

\vspace{-.4cm}

\begin{abstract}
A bounded operator $T$ on a separable, complex Hilbert space is said to be odd symmetric if $I^*T^tI=T$ where $I$ is a real unitary satisfying $I^2=-\one$ and $T^t$ denotes the transpose of $T$. It is proved that such an operator can always be factorized as $T=I^*A^tIA$ with some operator $A$. This generalizes a result of Hua and Siegel for matrices. As application it is proved that the set of odd symmetric Fredholm operators has two connected components labelled by a $\ZM_2$-index given by the parity of the dimension of the kernel of $T$. This recovers a result of Atiyah and Singer. Two examples of $\ZM_2$-valued index theorems are provided, one being a version of the Noether-Gohberg-Krein theorem with symmetries and the other an application to topological insulators.
\end{abstract}

\vspace{-.5cm}

\section{Resum\'e}
\label{sec-Z2ind}

Let $\Hh$ be a separable, complex Hilbert space equipped with a complex conjugation $\Cc$, namely an anti-linear involution. For $T$ from the set $\BM(\Hh)$ of bounded linear operators, one can then dispose of its complex conjugate $\overline{T}=\Cc T\Cc$ and its transpose $T^t=(\overline{T})^*$. Then $T$ is called real if $T=\overline{T}$ and symmetric if $T=T^t$. Let now further be given a real skew-adjoint unitary operator $I$ on $\Hh$.  Skew-adjointness $I^*=-I$ of $I$ is equivalent to $I^2=-\one$ and implies that the spectrum of $I$ is $\{-\imath,\imath\}$. Such an $I$ exists if and only if $\Hh$ is even or infinite dimensional. One may assume $I$ to be in the normal form $I=\left(\begin{smallmatrix} 0 & -\one \\ \one & \;\;0 \end{smallmatrix}\right)$, see Proposition~\ref{prop-standardforms} below. This paper is about bounded linear operators $T\in\BM(\Hh)$ which are odd symmetric w.r.t. $I$ in the sense that
\begin{equation}
\label{eq-oddsym}
I^*\,\overline{T}\,I\;=\;T^*
\qquad
\mbox{\rm or equivalently}
\qquad
I^*\,T^t\,I\;=\;T
\;.
\end{equation}
The set of bounded odd symmetric operators is denoted by $\BM(\Hh,I)$. Condition \eqref{eq-oddsym} looks like a quaternionic condition, but actually a quaternionic operator rather satisfies $I^*\overline{T}I=T$ and the set of quaternionic operators forms a multiplicative group, while $\BM(\Hh,I)$ does not. There was some activity on odd symmetric operators in the russian literature (where a different terminology was used), as is well-documented in \cite{Zag}, but not on the main questions addressed in this paper. The following elementary properties can easily be checked.


\begin{proposi}
\label{prop-algebra}
$\BM(\Hh,I)$ is a linear space and

\vspace{.1cm}

\noindent {\rm (i)} $T\in\BM(\Hh,I)$ if and only if $T^*\in\BM(\Hh,I)$.

\vspace{.1cm}

\noindent {\rm (ii)} If $T,T'\in\BM(\Hh,I)$ and $n\in\NM$, then $T^n\in\BM(\Hh,I)$ and $TT'+T'T\in\BM(\Hh,I)$.

\vspace{.1cm}

\noindent {\rm (iii)} For an invertible operator, $T\in\BM(\Hh,I)$ if and only if $T^{-1}\in\BM(\Hh,I)$.

\vspace{.1cm}

\noindent {\rm (iv)} For $A\in\BM(\Hh)$ and $T\in\BM(\Hh,I)$, one has $\frac{1}{2}(I^*A^tI+A)\in\BM(\Hh,I)$ and $I^*A^t T IA\in\BM(\Hh,I)$.

\vspace{.1cm}

\noindent {\rm (v)}  $T\in\BM(\Hh,I)$ if and only if $B=IT$ {\rm (}or $B=TI${\rm )} is skew-symmetric, namely $B^t=-B$.

\vspace{.1cm}

\noindent {\rm (vi)}
If the polar decomposition of $T\in\BM(\Hh,I)$ is $T=V|T|$ where $V$ is the unique partial 

isometry with $\Ker(T)=\Ker(V)$, then the polar decomposition of $T^*$ is $T^*=I^*\overline{V}I|T^*|$ 

and $|T^*|=I^*\overline{|T|}I$.

\end{proposi}

The factorization property stated in (iv) characterizes odd symmetric operators as is shown in the first main result of this paper stated next: 

\begin{theo}
\label{theo-factorizing} 
Any $T\in\BM(\Hh,I)$ is of the form $T=I^*A^tIA$ for some $A\in\BM(\Hh)$. If $\Ker(T)$ is either even dimensional or infinite dimensional, one moreover has $\Ker(A)=\Ker(T)$.
\end{theo}

For finite dimensional $\Hh$ this is due to \cite{Hua} (who stated a decomposition for skew-symmetric matrices which readily implies the above), but the proof presented below actually rather adapts the finite-dimensional argument of \cite[Lemma~1]{Sie}. 
Before going into the proof in Section~\ref{sec-proof}, let us give a summary of the remainder of the paper, consisting mainly of spectral-theoretic applications which are ultimately based on Theorem~\ref{theo-factorizing}. Item (v) of Propostion~\ref{prop-algebra} shows that there is a direct connection between odd symmetric and skew-symmetric operators. Hence one may expect that there is nothing interesting to be found in the spectral theory of odd symmetric operators in the case of a finite dimensional Hilbert space, but in fact these matrices have even multiplicities (geometric, algebraic, actually every level of the Jordan hierarchy). Actually, the spectra of $T$ and $B=IT$ have little in common as $T\psi=\lambda\psi$ is equivalent to $(B-\lambda I)\psi=0$. For $k\geq 1$ and $\lambda\in\CM$, let $d_k(T,\lambda)$ denote the dimension of the kernel of $(T-\lambda\one)^k$.

\begin{proposi}
\label{prop-finitecase} 
Let $T\in\BM(\Hh,I)$ where $\Hh$ is finite dimensional. Then $d_k(T,\lambda)$ and $d_1(T^*T,\lambda)$ are even for all $\lambda\in\CM$.
\end{proposi}

In the case of a self-adjoint or unitary odd symmetric operator $T$, this degeneracy is known as Kramers degeneracy \cite{Kra} and possibly the first trace of this in the mathematics literature is \cite[Theorem~6]{Hua}. The author could not localize any reference for the general fact of Proposition~\ref{prop-finitecase}, but after producing various proofs he realized that there is a simple argument basically due to \cite{Hua} and appealing to the Pfaffian.  A crucial difference between the self-adjoint and general case is that the generalized eigenspaces need not be invariant under $I$ in the latter case. Let us also point out that $T^*T$ is {\it not} odd symmetric, but nevertheless has even degeneracies. By an approximation argument, the even degeneracy extends to the set $\KM(\Hh,I)$ of compact odd symmetric operators.

\begin{proposi}
\label{prop-specialcase} 
Let $K\in\KM(\Hh,I)$ and $\lambda\not = 0$. Then $d_k(K,\lambda)$ is even for all $k\geq 1$.
\end{proposi}

The next result of the paper is about the subset $\FM(\Hh,I)$ of bounded odd symmetric Fredholm operators furnished with the operator norm topology. Recall that $T\in\BM(\Hh)$ is a Fredholm operator if and only if kernel $\Ker(T)$ and cokernel $\Ker(T^*)$ are finite dimensional and the range of $T$ is closed. Then the Noether index defined as $\Ind(T)=\dim(\Ker(T))-\dim(\Ker(T^*))$ is a compactly stable homotopy invariant. For an odd symmetric Fredholm operator, one has $\Ker(T^*)=I\Cc\Ker(T)$ so that the Noether index  vanishes. Nevertheless, there is an interesting invariant given by the parity of the dimension of the kernel which is sometimes also called the nullity.

\begin{theo}
\label{theo-Z2} 
Let $T\in \FM(\Hh,I)$ and $K\in\KM(\Hh,I)$. Set $\Ind_2(T)=\dim(\Ker(T))\,\mbox{\rm mod}\,2\in\ZM_2$.

\vspace{.1cm}

\noindent {\rm (i)} If  $ \Ind_2(T)=0$, there exists a finite-dimensional odd symmetric partial isometry $V\in\BM(\Hh,I)$ 

such that $T+V$ is invertible. 


\vspace{.1cm}

\noindent {\rm (ii)} $\Ind_2(T+K)=\Ind_2(T)$

\vspace{.1cm}

\noindent {\rm (iii)} The map $T\in \FM(\Hh,I)\mapsto \Ind_2(T)$ is continuous. 

\vspace{.1cm}

\noindent {\rm (iv)} $\FM(\Hh,I)$ is the disjoint union of two open and connected components $\FM_0(\Hh,I)$ and $\FM_1(\Hh,I)$ 

labelled by $\Ind_2$. 

\end{theo}

This theorem is not new as it can be deduced from the paper of Atiyah and Singer \cite{AS} because $\FM(\Hh,I)$ can be shown to be isomorphic to the classifying space $\Ff^2(\Hh_\RM)$ defined  in \cite{AS}. This isomorphism will be explained in detail following the proof of Theorem~\ref{theo-Z2} in Section~\ref{sec-proof2}. Nevertheless, even given this connection, the proof in \cite{AS} is quite involved. Here a detailed and purely functional analytic argument based on the factorization property in Theorem~\ref{theo-factorizing} and the Kramers degeneracy in Proposition~\ref{prop-specialcase} is presented. 

\vspace{.2cm}

It is worth noting that Theorem~\ref{theo-Z2} can also be formulated for skew-symmetric operators by using the correspondence of Proposition~\ref{prop-algebra}(v), but the author feels that there are two good reasons not to do so: the spectral degeneracy is linked to odd symmetric rather than skew-symmetric operators, and in applications to time-reversal symmetric quantum mechanical systems (where $I$ is the the rotation in spin space for a system with half-integer spin) one is naturally lead to odd symmetric operators. The $\ZM_2$-index has a number of further basic properties, like $\Ind_2(T)=\Ind_2(T^*)$ and $\Ind_2(T\oplus T')=\Ind_2(T)+\Ind_2(T')\;{\rm mod}\,2$, but the author was not able to find a trace formula for the $\ZM_2$-index similar to the Calderon-Fedosov formula for the Noether index. Theorem~\ref{theo-Z2} is restricted to bounded Fredholm operators, but readily extends to unbounded operators with adequate modifications.

\vspace{.2cm}

Just as for Fredholm operators with non-vanishing Noether index, an example of an  odd symmetric operator with a non-trivial $\ZM_2$-index can be constructed from the shift operator $S$ on $\ell^2(\NM)$  defined as usual by $S|n\rangle=\delta_{n\geq 2}|n-1\rangle$: the operator $T=S\oplus S^*$ on $\ell^2(\NM)\otimes\CM^2$ is odd symmetric w.r.t. $I=\left(\begin{smallmatrix}  0 & -\one \\ \one & \;\; 0 \end{smallmatrix}\right)$ and has a one-dimensional kernel. 

\vspace{.2cm}

Building on this example, a $\ZM_2$-valued index theorem is proved in Section~\ref{sec-oddGK}. It considers the setting of the Noether-Gohberg-Krein index theorem connecting the winding number of a function $z\in\SM^1\mapsto f(z)$ on the unit circle to the index of the associated Toeplitz operator $T_f$. If the function is matrix-valued and has the symmetry property $I^*f(\overline{z})I=f(z)^t$, then the Toeplitz operator is odd symmetric and its $\ZM_2$-index is proved to be equal to an adequately defined $\ZM_2$-valued winding number of $f$, which can be seen as a topological index associated to $f$. It ought to be stressed that the examples of index theorems in \cite{AS2} invoked the classifying space $\Ff^1(\Hh_\RM)$ of skew-symmetric Fredholm operators on a real Hilbert space rather than $\Ff^2(\Hh_\RM)$. Hence they are of different nature. Both results can be described in the realm of $KR$-theory \cite{Sc}. The aim of our presentation here is not to give the most general version of such a $\ZM_2$-index theorem, but rather to provide a particularly simple example. As a second example, again using $\FM(\Hh,I)\cong \Ff^2(\Hh_\RM)$ and not $\Ff^1(\Hh_\RM)$, Section~\ref{sec-topins} considers two-dimensional topological insulators which have half-integer spin and time-reversal symmetry. In these systems a $\ZM_2$-index is defined and shown to be of physical importance, as it is shown to be equal to the parity of the spin Chern numbers. In another publication \cite{DS} (actually written after a first version of this work was available), the $\ZM_2$-index is also linked to a natural spectral flow in these systems.

\vspace{.2cm}

Before turning tot the proofs of Theorems~\ref{theo-factorizing} and \ref{theo-Z2} as well as details on the $\ZM_2$-index theorems, let us briefly consider quaternionic and even symmetric Fredholm operators in order to juxtapose them with odd symmetric Fredholm operators. The following result follows from a standard Kramers degeneracy argument.

\begin{proposi}
\label{prop-quaternionic} Let $T\in\BM(\Hh)$ be a quaternionic Fredholm operator, namely $I^*\overline{T}I=T$. Then $\Ind(T)\in 2\,\ZM$ is even.
\end{proposi}

Next suppose given a real unitary $J$ on $\Hh$ with $J^2=\one$. This implies $J^*=J=J^{-1}$ and that the spectrum of $J$ is contained in $\{-1,1\}$. Note that, in particular, $J=\one$ is also possible. Then an operator is called even symmetric w.r.t. $J$ if $JT^tJ=T$, which is completely analogous to \eqref{eq-oddsym}. Such operators were studied in \cite{GP,Zag} and the references cited therein, and a variety of different terminologies was used for them. Again Proposition~\ref{prop-algebra} remains valid for the set $\BM(\Hh,J)$ of even symmetric operators  except for item (v), the equivalent of which is that the operator $B=JT$ is symmetric $B^t=B$ if and only if $T\in \BM(\Hh,J)$. Next let us consider the set $\FM(\Hh,J)$ of even symmetric Fredholm operators. The following result, analogous to Theorems~\ref{theo-factorizing} and \ref{theo-Z2}, shows that there is no interesting topology in $\FM(\Hh,J)$.

\begin{theo}
\label{theo-evensym}
Let $J$ be a real unitary on $\Hh$ with $J^2=\one$. Then for any $T\in\BM(\Hh,J)$ there exists  $A\in\BM(\Hh)$ such that $T=JA^tJA$ and $\Ker(A)=\Ker(T)$. The set $\FM(\Hh,J)$ is connected. 
\end{theo}


\section{Proof of the factorization property}
\label{sec-proof}

In the following, a real unitary operator is called orthogonal. The following result was mentioned in the first paragraph of the paper. 

\begin{proposi}
\label{prop-standardforms}
Let $I$ and $J$ be real unitaries with $I^2=-\one$ and $J^2=\one$. Then there are orthogonal operators $O$ and $O'$ such that $O^tIO$ and $(O')^tJO'$ are of the normal form 
$$
O^tI O
\;=\;
\left(\begin{matrix} 0 & -\one \\ \one & \;\;0\end{matrix}\right)
\;,
\qquad
(O')^tJ O'
\;=\;
\left(\begin{matrix} \one & 0 \\ 0& -\,\one \end{matrix}\right)
\;.
$$
If $T$ is odd symmetric w.r.t. $I$, then $O^tTO$ is odd symmetric w.r.t. $O^tIO$. Similarly, if $T$ is even symmetric w.r.t. $J$, then $(O')^tTO'$ is even symmetric w.r.t. $(O')^tJO'$.
\end{proposi}

\noindent {\bf Proof.} Let us focus on the case of $I$. The spectrum of $I$ is $\{\imath,-\imath\}$ and the eigenspaces $\Ee_{-\imath}$ and $\Ee_{\imath}$ are complex conjugates of each other and are, in particular, of same dimension. Hence there is a unitary $V=(\overline{v},v)$ built from an orthonormal basis $v=(v_1,v_2,\ldots)$ of $\Ee_\imath$ such that $V^*I V=-\imath\,\left(\begin{smallmatrix} \one & \;\;0 \\ 0 & -\one \end{smallmatrix}\right)$. Now the Cayley transform $C$ achieves the following
\begin{equation}
\label{eq-Cayley}
C^*\,
\begin{pmatrix}
\one & 0 \\ 0 & -\one
\end{pmatrix}
\,C\;=\;\imath\,
\begin{pmatrix}
0 & -\,\one \\ \one & 0
\end{pmatrix}
\;,
\qquad
C
\;=\;
\frac{1}{\sqrt{2}}\,
\begin{pmatrix}
\one & -\imath\,\one \\ \one & \imath\,\one
\end{pmatrix}
\;.
\end{equation}
Hence $O=VC$ is both real and satisfies the desired equality. The reality of $O$ also implies the  claim about odd symmetric operators.
\hfill $\Box$

\vspace{.2cm}

As a preparation for the proof of Theorem~\ref{theo-factorizing}, let us begin with the following result of independent interest. A related result in finite dimension was proved in \cite{Hua}, but the argument presented here adapts the proof of Lemma~1 in \cite{Sie} to the infinite dimensional situation. A preliminary result to Proposition~\ref{prop-normalskew} can be found in \cite{LZ}.

\begin{proposi} 
\label{prop-normalskew}
Let $N\in\BM(\Hh)$ be a normal and skew-symmetric operator on a complex Hilbert space $\Hh$ with complex conjugation $\Cc$. Then there exists an orthogonal operator $O:\Hh'\to\Hh$ from a complex Hilbert space $\Hh'$ onto $\Hh$ and a bounded operator $M$ with trivial kernel such that in an adequate grading of $\Hh'$
\begin{equation}
\label{eq-orthfact}
O^tNO
\;=\;
\begin{pmatrix}
M & 0 & 0 \\ 0 & M & 0 \\ 0 & 0 & 0
\end{pmatrix}
\begin{pmatrix}
0 & -\,\one & 0
\\
\one & 0 & 0
\\
0 & 0 & 0
\end{pmatrix}
\begin{pmatrix}
M & 0 & 0 \\ 0 & M & 0 \\ 0 & 0 & 0
\end{pmatrix}^t
\;.
\end{equation}
\end{proposi}

\noindent {\bf Proof.}  By normality, $\Ker(N)=\Ker(N^*)$, and skew symmetry $\Ker(N^*)=\Cc\Ker(N)$. Thus $\Ker(N)=\Cc \Ker(N)$ is invariant under complex conjugation $\Cc$. It is possible to choose a real orthonormal basis of $\Ker(N)$. This is used as the lowest block of $O$ in \eqref{eq-orthfact} corresponding to the kernel of $N$. Now one can restricted $N$ to $\Ran(N)=\Ker(N)^\perp$ which is also a closed subspace that is invariant under $\Cc$.  Equivalently, it is possible to focus on the case where  $\Ker(N)=\{0\}$. Recall that the complex conjugate and transpose are defined by $\overline{N}=\Cc N\Cc$ and $N^t=\Cc N^*\Cc$ and skew-symmetry means $N^t=-N$. Then by normality
$$
N^*N\;=\;-\,\overline{N}\,N\;=\;-\,N\,\overline{N}
\;,
$$
so that $N^*N$ is a real operator. Let us decompose
$$
N
\;=\;
N_1\,+\,\imath\,N_2
\;,
\qquad
N_1\,=\,\frac{1}{2}(N-\overline{N})\;,
\;\;\;
N_2\,=\,\frac{1}{2\,\imath}(N+\overline{N})
\;.
$$
Then $N_1$ and $N_2$ are purely imaginary, self-adjoint and commute due to the reality of 
$$
N\,\overline{N}
\;=\;
-\,(N_1)^2\,-\,(N_2)^2\,+\,\imath(N_1N_2-N_2N_1)
\;.
$$
Thus they can and will be simultaneously diagonalized. Also, one has $\Ker(N_1)\cap\Ker(N_2)=\{0\}$ because otherwise $N$ would have a non-trivial kernel. Furthermore, the skew-symmetry of $N_j$, $j=1,2$, implies that the spectrum satisfies $\sigma(N_j)=-\sigma(N_j)$ and the spectral projections $P_j(\Delta)$ satisfy
\begin{equation}
\label{eq-projsym}
\overline{P_j(\Delta)}
\;=\;
P_j(-\Delta)
\;,
\qquad
\Delta\subset \RM
\;.
\end{equation}
In fact, for any $n\in\NM$ and $\alpha\in\CM$, one has $\overline{\alpha\, N_j^n}=\overline{\alpha}\,(-N_j)^n$ and hence for any continuous function $f:\RM\to\CM$ also $\overline{f(N_j)}=\overline{f}(-N_j)$. By spectral calculus, this implies \eqref{eq-projsym}. Next let us set $\Ee_\pm=\Ran(P_1(\RM_\pm))$ where $\RM_+=(0,\infty)$ and $\RM_-=(-\infty,0)$, as well as $\Ee_0=\ker(N_1)$. Then $\overline{\Ee_+}=\Ee_-$ and  $\overline{\Ee_0}=\Ee_0$ and $\Hh=\Ee_+\oplus\Ee_-\oplus\Ee_0$. Now let us apply the spectral theorem to $N_{1,+}=N_1|_{\Ee_+}$ which has its spectrum in $\RM_+$. It furnishes a sequence of measures $\mu_n$ and a unitary $u:\bigoplus_{n\geq 1}L^2(\RM_+,\mu_n)\to\Ee_+$ such that
$$
u^*\,N_{1,+}\,u
\;=\;
M_{1,+}
\;,
$$
where $M_{1,+}:\bigoplus_{n\geq 1}\,L^2(\RM_+,\mu_n)\to \bigoplus_{n\geq 1}\,L^2(\RM_+,\mu_n)$ is the real multiplication operator given by $(M_{1,+}\psi)(x)=x\,\psi(x)$. Due to  \eqref{eq-projsym} and because $N_1$ is purely imaginary,  $\overline{u}=\Cc\, u\,\Cc:\bigoplus_{n\geq 1}\,L^2(\RM_+,\mu_n)\to\Ee_-$ leads for $N_{1,-}=N_1|_{\Ee_-}$ to
$$
\overline{u}^*\,N_{1,-}\,\overline{u}
\;=\;
-\,M_{1,+}
\;.
$$
Taking the direct sum of $u$ and $\overline{u}$, one obtains a unitary
$$
\begin{pmatrix}
u & 0 \\ 0 & \overline{u}
\end{pmatrix}
\;:\;
\bigoplus_{n\geq 1}L^2(\RM_+,\mu_n)\otimes\CM^2
\;\to\;
\Ee_+\oplus\Ee_-
\;,
$$
such that
$$
\begin{pmatrix}
u & 0 \\ 0 & \overline{u}
\end{pmatrix}^*
\,N_1\,
\begin{pmatrix}
u & 0 \\ 0 & \overline{u}
\end{pmatrix}
\;=\;
\begin{pmatrix}
M_{1,+} & 0
\\
0 & -\,M_{1,+}
\end{pmatrix}
\;.
$$
As $N_1$ and $N_2$ commute, $u$ can furthermore be chosen such that
$$
\begin{pmatrix}
u & 0 \\ 0 & \overline{u}
\end{pmatrix}^*
\,N_2\,
\begin{pmatrix}
u & 0 \\ 0 & \overline{u}
\end{pmatrix}
\;=\;
\begin{pmatrix}
M_{2,+} & 0
\\
0 & -\,M_{2,+}
\end{pmatrix}
\;.
$$
where $M_{2,+}=u N_2|_{\Ee_+}u^*$ is also a multiplication operator which is, however, not positive, and it was used that $N_2$ is purely imaginary. Furthermore, $\Ee_0$ is a real subspace that is invariant under $N_2$ and $\Ker(N_2|_{\Ee_0})=\{0\}$. Following the above argument, now for $N_2$, one can decompose $\Ee_0=\Ee_{0,+}\oplus\Ee_{0,-}$ in the positive and negative subspace of $N_2$ and obtains a sequence of measures $\nu_n$ on $\RM_+$ and a unitary $v:\bigoplus_{n\geq 1}L^2(\RM_+,\nu_n)\to\Ee_{0,+}$ such that
$$
\begin{pmatrix}
v & 0 \\ 0 & \overline{v}
\end{pmatrix}^*
\,N_2\,
\begin{pmatrix}
v & 0 \\ 0 & \overline{v}
\end{pmatrix}
\;=\;
\begin{pmatrix}
M_{0,+} & 0
\\
0 & -\,M_{0,+}
\end{pmatrix}
\;.
$$
Combining and rearranging, this provides a spectral representation for $N=N_1+\imath N_2$:
$$
U^*\,N\,U
\;=\;
\begin{pmatrix}
M_{1,+}\,+\,\imath\,M_{2,+} & 0 & 0 & 0
\\
0 & \imath\,M_{0,+} & 0 & 0
\\
0& 0 & -\,(M_{1,+}\,+\,\imath\,M_{2,+}) & 0
\\
0 & 0 & 0 & -\,\imath\,M_{0,+}
\end{pmatrix}
\;,
$$
where $U=u\oplus v\oplus\overline{u}\oplus\overline{v}$.  Now let us conjugate this equation with the Cayley transformation defined in \eqref{eq-Cayley} where each entry corresponds to $2\times 2$ blocks. Then one readily checks that $O=C^*UC$ is  an orthogonal operator and 
$$
O^t\,N\,O
\;=\;
\begin{pmatrix}
0 & 0 & -\,\imath(M_{1,+}\,+\,\imath\,M_{2,+}) & 0
\\
0 & 0 & 0 & -\,\imath\,M_{0,+}
\\
\imath(M_{1,+}\,+\,\imath\,M_{2,+}) & 0 & 0 & 0
\\
0 & \imath\,M_{0,+} & 0 & 0
\end{pmatrix}
\;.
$$
Now all the operators on the r.h.s. are diagonal multiplication operators and one may set
$$
M
\;=\;
\begin{pmatrix}
\imath M_{1,+}-M_{2,+} & 0 
\\
0 & \imath\,M_{0,+}
\end{pmatrix}^{\frac{1}{2}}
\;.
$$
This leads to \eqref{eq-orthfact} in the case with trivial kernel, in the grading $\Hh_+\oplus\Hh_-$ where $\Hh_\pm=\{\psi\pm\overline{\psi}\,|\,\psi\in\Ee_+\oplus\Ee_{0,+}\}$. How to include the kernel of $N$ was already explained above.
\hfill $\Box$

\begin{proposi} 
\label{prop-skew}
Let $B\in\BM(\Hh)$ be a skew-symmetric operator on a complex Hilbert space $\Hh$ with complex conjugation $\Cc$. Then there exists a unitary operator $U:\Hh'\to\Hh$ from a complex Hilbert space $\Hh'$ onto $\Hh$ and a bounded operator $M$ with trivial kernel such that in an adequate grading of $\Hh'$
\begin{equation}
\label{eq-unitfact}
U^tBU
\;=\;
\begin{pmatrix}
M & 0 & 0 \\ 0 & M & 0 \\ 0 & 0 & 0
\end{pmatrix}
\begin{pmatrix}
0 & -\,\one & 0
\\
\one & 0 & 0
\\
0 & 0 & 0
\end{pmatrix}
\begin{pmatrix}
M & 0 & 0 \\ 0 & M & 0 \\ 0 & 0 & 0
\end{pmatrix}^t
\;.
\end{equation}
\end{proposi}

\noindent {\bf Proof.}  
By the spectral theorem, there exist measures $\mu_n$ on $\RM_\geq =[0,\infty)$ and a unitary $W:\Hh\to \bigoplus_{n\geq 1}\,L^2(\RM_\geq,\mu_n)$ such that
$$
B^*B\;=\;W^*DW
\;,
$$
where $D:\bigoplus_{n\geq 1}\,L^2(\RM_\geq,\mu_n)\to \bigoplus_{n\geq 1}\,L^2(\RM_\geq,\mu_n)$ is the multiplication operator $(D\psi)(x)=x\,\psi(x)$. Now 
$$
BB^*\;=\;W^tD\overline{W}
\;.
$$
Let us set $N=\overline{W}BW^*$. Then $N$ is skew-symmetric and normal because $N^*N=D=NN^*$. Hence Proposition~\ref{prop-normalskew} can be applied. Setting $U=W^*O$ concludes the proof.
\hfill $\Box$



\begin{proposi} 
\label{prop-unitskew2}
Let $B\in\BM(\Hh)$ be a skew-symmetric operator on a complex Hilbert space $\Hh$ with complex conjugation $\Cc$. Suppose that $\dim(\Ker(B))$ is even or infinite. Let $I$ be a unitary with $I^2=-\one$. Then there exists an operator $A\in\BM(\Hh)$ with $\Ker(A)=\Ker(B)$ such that
$$
B
\;=\;
A^tIA
\;.
$$
\end{proposi}

\noindent {\bf Proof.} If $\dim(\Ker(N))$ is even or infinite one can modify \eqref{eq-unitfact} to
$$
U^tBU
\;=\;
\begin{pmatrix}
M & 0 & 0 & 0\\ 
0 & 0 & 0 & 0 \\
0 & 0 & M & 0 \\ 
0 & 0 & 0 & 0
\end{pmatrix}
\begin{pmatrix}
0 & 0 & -\,\one & 0
\\
0 & 0 & 0 & -\,\one
\\
\one & 0 & 0 & 0
\\
0 & \one & 0 & 0
\end{pmatrix}
\begin{pmatrix}
M & 0 & 0 & 0\\ 
0 & 0 & 0 & 0 \\
0 & 0 & M & 0 \\ 
0 & 0 & 0 & 0
\end{pmatrix}^t
\;.
$$
Inserting adequate orthogonals provided by Proposition~\ref{prop-standardforms}, the operator in the middle becomes $I$ as claimed.
\hfill $\Box$

\vspace{.2cm}

\noindent {\bf Proof} of Theorem~\ref{theo-factorizing}.  
Associated to $T\in\BM(\Hh,I)$ is the skew-symmetric $B=IT$. Applying Proposition~\ref{prop-unitskew2} to $B$ provides the desired factorization of $T$ for the case of an even dimension or an infinite dimensional kernel. For the odd dimensional case, let us choose a real orthonormal basis and let $C$ be the associated unilateral shift, namely a real partial isometry with $CC^t=\one$ and $\one-C^tC$ an orthogonal projection of dimension $1$. Now the operator $I^*C^tTIC$ is odd symmetric by Proposition~\ref{prop-algebra}(iv) and its kernel is even dimensional because $C^t$ has trivial kernel and the range of $C$ is all $\Hh$. By the above, $I^*C^tTIC=IA^tIA$ for some $A\in\Bb(\Hh)$. Thus $T=I^*(AC^tI)^tI(AC^tI)$.
\hfill $\Box$

\vspace{.2cm}

\section{Proof of properties of the $\ZM_2$ index}
\label{sec-proof2}

\noindent {\bf Proof} of Proposition~\ref{prop-finitecase}. Following \cite{Hua}, let us first prove that the spectrum of the non-negative operator $T^*T$ has even degeneracy. If $T$ has a kernel, choose a small $\epsilon$ such that $T+\epsilon\,\one$ has a trivial kernel. Then $B=I(T+\epsilon\,\one)$ is skew-symmetric and invertible. One has $\det(B^*B-\lambda \one)=\det(B)\det(B^*-\lambda\,B^{-1})$. As $B^*-\lambda B^{-1}$ is skew-symmetric, its determinant is the square of the Pfaffian and thus, in particular, has roots of even multiplicity. Consequently the spectrum of $B^*B=(T+\epsilon\,\one)^*(T+\epsilon\,\one)$ has even multiplicities. Taking $\epsilon\to 0$ shows that also $T^*T$ has even multiplicities, namely $d_1(T^*T,\lambda)$ is even. Now $\Ker(T)=\Ker(T^*T)$ so that also $d_1(T,0)=d_1(T^*T,0)$ is even. Further, as $T^k$ is also odd symmetric by Proposition~\ref{prop-algebra}, also $d_k(T,0)=d_1(T^k,0)$ is even.  For any other eigenvalue $\lambda$, one uses the odd symmetric matrix $T-\lambda \one$ to deduce that $d_k(T,\lambda)=d_k(T-\lambda\one,0)$ is also even.
\hfill $\Box$

\vspace{.2cm}

\noindent {\bf Proof} of Proposition~\ref{prop-specialcase}. Let $R_{n}$ be a sequence of $2n$-dimensional real projections commuting with $I$ and converging weakly to $\one$. The existence of such a sequence can readily be deduced from Proposition~\ref{prop-standardforms}. Set $K_n=R_{n} KR_{n}$. Then $K_n$ restricted to the range of $R_n$ is a finite dimensional odd symmetric operator which has even degeneracies by Proposition~\ref{prop-finitecase}. Let us set $T_n=(K_n-\lambda\one)^k$. Then the spectrum of $T_n^*T_n$ consists of the infinitely degenerate point $|\lambda|^{2k}$ and a finite number of positive eigenvalues which have even degeneracies. Now $T_n$ converges to $T=(K-\lambda\one)^k$ in the norm topology. Thus the eigenvalues of $T_n^*T_n$ and associated Riesz projections converge the eigenvalues and Riesz projections of $T^*T$ \cite[VIII.1]{Kat}. As all eigenvalues of $T_n^*T_n$ have even degeneracy for all $n$, it follows that, in particular, the kernel of $T^*T$ also has even degeneracy. But $\Ker(T)=\Ker(T^*T)$ and $\dim(\Ker(T))=d_k(K,\lambda)$ completing the proof.
\hfill $\Box$

\vspace{.2cm}

\noindent {\bf Proof} of the Theorem~\ref{theo-Z2}(i). Because the Noether index vanishes, one has $\dim(\Ker(T))=\dim(\Ker(T^*))<\infty$. By hypothesis, $\dim(\Ker(T))$ is even, say equal to $2N$. Let $(\phi_n)_{n=1,\ldots,2N}$ be an orthonormal basis of $\Ker(T)$. As \eqref{eq-oddsym} implies $\Ker(T^*)=I\Cc\Ker(T)$, an orthonormal basis of $\Ker(T^*)$ is given by $(I\,\overline{\phi}_n)_{n=1,\ldots,2N}$. Using Dirac's Bra-Ket notations, let us introduce
\begin{equation}
\label{eq-Vdef}
V
\;=\;
\sum_{n=1}^N
\left(
I\,|\overline{\phi}_n\rangle\langle\phi_{n+N}|
\,-\,
I\,|\overline{\phi}_{n+N}\rangle\langle\phi_{n}|
\right)
\;.
\end{equation}
Then $V^*V$ and $VV^*$ are the projections on $\Ker(T)$ and $\Ker(T^*)$, and one has indeed $I^*V^tI=V$. From now on the proof follows standard arguments. To check injectivity of $T+V$, let $\psi\in\Hh$ satisfy $(T+V)\psi=0$. Then 
$$
T\psi\; =\; -V\psi\in T\Hh\, \cap\, \Ran(V)
\; =\; T\Hh\, \cap\, \Ker(T^*)
\;=\; 
T\Hh\, \cap\, \Ran(T)^\perp
\; =\; 
\{0\}
\; ,
$$
so that  $T\psi=0$ und $V^*V\psi=0$ and $\psi\in\Ker(T)\cap\Ker(V^*V)=\Ker(T)\cap
\Ker(T)^\perp=\{0\}$. Furthermore, $T+V$ is surjective, because $V\Ker(T)=\Ker(T^*)$ implies
$$
(T+V)(\Hh)\; =\; (T+V)\big(\Ker(T)^\perp\oplus\Ker(T)\big)\; =\; T(\Hh)\oplus\Ker(T^*)\; =\; 
T\Hh\oplus(T\Hh)^\perp\; =\; \Hh
\; ,
$$
where the last equality holds because the range of the Fredholm operator $T$ is closed. Hence $T+V$ is bijective and bounded, so that the Inverse Mapping Theorem implies that it is also has a bounded inverse.
\hfill $\Box$

\vspace{.2cm}

\noindent {\bf Proof} of the Theorem~\ref{theo-Z2}(ii). Let us first suppose that $\Ind_2(T)=0$. By Theorem~\ref{theo-Z2}(i) there is a finite-dimensional odd symmetric partial isometry such that $T+V$ is invertible. According to Theorem~\ref{theo-factorizing} there exists an invertible operator $A\in\BM(\Hh)$ such that $T+V=I^*A^tIA$. Thus
$$
T+K
\;=\;
I^*A^t \left(\one+(A^t)^{-1}I(K-V)IA^{-1}I^*\right)IA
\;.
$$
Now $K'=(A^t)^{-1}I(K-V)IA^{-1}$ is compact and by Proposition~\ref{prop-specialcase}  the dimension of the kernel of $\one+K'$ is even dimensional. This dimension is not changed by multiplication with invertible operators. Now let $\Ind_2(T)=1$. Let $C$ be a Fredholm operator with $1$-dimensional kernel and trivial cokernel and set 
\begin{equation}
\label{eq-Thelp}
\widehat{T}
\;=\;
I^*C^tTIC
\;.
\end{equation}
Then $\widehat{T}$ is odd symmetric by Proposition~\ref{prop-algebra}(iv) and its kernel is even dimensional because $C^t=(\overline{C})^*$ has trivial kernel and the kernel of $T$ lies in the range of $C$. Consequently, $\Ind_2(\widehat{T})=0$  and the compact stability of its index is already guaranteed. Thus $\widehat{T+K}=I^*C^t(T+K)IC$ has vanishing $\ZM_2$ index and thus even dimensional kernel. One concludes that $T+K$ has odd dimensional kernel so that $\Ind_2(T+K)=1$.
\hfill $\Box$

\vspace{.2cm}

\noindent {\bf Proof} of the Theorem~\ref{theo-Z2}(iii) and (iv). Actually (iii) follows once it is proved that the sets $\FM_0(\Hh,I)$ and $\FM_1(\Hh,I)$ of odd symmetric Fredholm operators with even and odd dimensional kernel are open in the operator topology. Let us first prove that $\FM_0(\Hh,I)$ is open. Let $T\in\FM_0(\Hh,I)$ and let $T_n\in\BM(\Hh,I)$ be a sequence of odd symmetric operators converging to $T$. By (i), there exists a finite dimensional partial isometry $V\in\BM(\Hh,I)$ such that $T+V$ is invertible. Thus
$$
T_n+V
\; =\; 
T+V+T_n-T
\; =\; 
(T+V)(\one+(T+V)^{-1}(T_n-T))
\;.
$$
For $n$ sufficiently large, the norm of $(T+V)^{-1}(T_n-T)$ is smaller than $1$, so that the Neumann series for the inverse of $\one+(T+V)^{-1}(T_n-T)$ converges. Hence $T_n+V$ is invertible and $\Ind_2(T_n)=0$ by (ii), namely $T_n\in \FM_0(\Hh,I)$ for $n$ sufficiently large. For the proof that also $\FM_1(\Hh,I)$ is open, let now $T\in\FM_1(\Hh,I)$ and $T_n\in\BM(\Hh,I)$ with $T_n\to T$ in norm. Then consider the operators $\widehat{T}$ and $\widehat{T_n}$ constructed as in \eqref{eq-Thelp}. They have vanishing $\ZM_2$-index so that the above argument applies again. 
%
%
It remains to show that $\FM_0(\Hh,I)$ and $\FM_1(\Hh,I)$ are connected. If $T\in\FM_0(\Hh,I)$, let again $V\in\BM(\Hh,I)$ be the finite dimensional partial isometry such that $T+V$ is invertible. Then $s\in[0,1]\mapsto T_s=T+sV$ is a path from $T$ to an invertible operator $T_1\in\FM_0(\Hh,I)$. Using Theorem~\ref{theo-factorizing} let us choose an invertible $A\in\BM(\Hh)$ such that $T_1=I^*A^tIA$. Because $A$ is invertible, the polar decomposition is of the form $A=e^{\imath H}|A|$ with a self-adjoint operator $H$ (so that the phase is a unitary operator). Thus $s\in[1,2]\mapsto A_s=e^{\imath H(2-s)}|A|^{2-s}$ is a norm continuous path of invertible operators from $A_1=A$ to $A_2=\one$. This induces the path $s\in[1,2]\mapsto T_s=I^*(A_s)^tIA_s\in\FM_0(\Hh,I)$ from $T_1$ to $T_2=\one$. This shows that $\FM_0(\Hh,I)$ is path connected. For the proof that also $\FM_1(\Hh,I)$ is path connected, one can use again $\widehat{T}$ with trivial index defined in \eqref{eq-Thelp}. Let us also assume that $C$ is a real partial isometry (such as a unilateral shift associated to a real orthonormal basis) so that $CC^*=CC^t=\one$.  By the above, there is a path $s\in[0,1]\mapsto \widehat{T}_s\in\FM_0(\Hh,I)$ from $\widehat{T}_0=\widehat{T}$ to $\widehat{T}_1=\one$. Then $s\in[0,1]\mapsto CI\widehat{T}_sC^tI^*$ is a path in $\FM_1(\Hh,I)$ from $T$ to $CIC^tI^*\in\FM_1(\Hh,I)$. As this hold for any $T\in\FM_1(\Hh,I)$, the proof is complete.
\hfill $\Box$

\vspace{.2cm}

Let us now explain in detail the connetion of $\FM(\Hh,I)$ to the classifying space $\Ff^2(\Hh_\RM)$ as defined in \cite{AS}. Atiyah and Singer consider a real Hilbert space $\Hh_\RM$ on which is given a linear operator $J$ satisfying $J^*=-J$ and $J^2=-\one$. Then $ \Ff^2(\Hh_\RM)$ is defined as the set of skew-adjoint Fredholm operators $A$ on $\Hh_\RM$ satisfying $AJ=-JA$. It is then shown to have exactly two connected components. To establish a ($\RM$-linear) bijection from $\Ff^2(\Hh_\RM)$ to $\FM(\Hh,I)$, let us choose a basis of $\Hh_\RM$ such that $J=\binom{0\;-\one}{\one\;\;\;0}$. In this basis, write $x=\binom{u}{v}\in \Hh_\RM$ and define $\varphi$ from $\Hh_\RM$ to a new vector space $\Hh$ by $\varphi(x)=u+\imath v$ where $\imath$ is the imaginary unit. Defining a scalar multiplication by complex scalars $\lambda=\lambda_\Re +\imath \lambda_\Im$ in the usual way by $\lambda(u+\imath v)=(\lambda_\Re u-\lambda_\Im v) +\imath(\lambda_\Re v+\lambda_\Im u)$, the vector space $\Hh$ becomes complex. This means that $J$ implements multiplication by $\imath$, namely $\imath \varphi(x)=\varphi(J x)$. Now let us introduce a scalar product and complex conjugation $\Cc$ on $\Hh$ by setting
$$
\langle \varphi(x)| \varphi(y)\rangle_{\Hh}
\;=\;
\langle x| y\rangle_{\Hh_\RM}
\;,
\qquad
\Cc \varphi(x)\;=\;u-\imath v \;\;\;\mbox{ for }x=\binom{u}{v}
\;.
$$
Resuming, via $\varphi$ the real Hilbert space $\Hh_\RM$ with skew-adjoint unitary $J$ can be seen as a complex Hilbert space $\Hh$ with a complex conjugation (or alternatively a real structure). Now given a linear operator $A$ on $\Hh_\RM$ satisfying $A=JAJ$, the operator $B=\varphi A\varphi^{-1}\Cc$ can be checked to be a $\CM$-linear operator on $\Hh$. Furthermore, the skew-adjointness of $A$ implies that $B=-B^t$. Explicitly, with the above identifications, if $A=\binom{a\;\;b}{b\;-a}$ with linear operators $a$ and $b$, then $B=a+\imath b$. Now the kernel of $A$ is invariant under $J$, and therefore $\Ker(B)=\Cc\varphi(\Ker(A))$ is a $\CM$-linear subspace with $\dim_\CM(\Ker(B))=\frac{1}{2}\dim_\RM(\Ker(A))$. As similar statements hold for the cokernels, one deduces in particular that $A$ is Fredholm on $\Hh_\RM$ if and only if $B$ is Fredholm on $\Hh$. Therefore $A\in\Ff^2(\Hh_\RM)\mapsto T=IB\in\FM(\Hh,I)$ is a bijection as claimed.

\vspace{.2cm}

Before going on to presenting $\ZM_2$-valued index theorems in the next sections, let us prove the remaining statements from the introduction.

\vspace{.2cm}

\noindent {\bf Proof} of the Proposition~\ref{prop-quaternionic}. As $T$ is quaterionic if and only if $T^*$ is quaternionic, it is sufficient to show that $\Vv=\Ker(T)$ is even dimensional. From $I^*\overline{T}I=T$ one infers $I\Cc \Vv=\Vv$. Actually any finite dimensional complex vector space with this property is even dimensional. Indeed, choose a non-vanishing $\phi_1\in\Vv$. Then $I\overline{\phi_1}\in \Vv$ and $\phi_1$ are linearly independent because $\phi_1=\lambda I\overline{\phi_1}$ for some $\lambda\in\CM$ leads to the contradiction $\phi_1=|\lambda|^2I^2\phi_1=-|\lambda|^2\phi_1$. Next choose $\phi_2$ in the orthogonal complement of the span of $\phi_1,I\overline{\phi_1}$. One readily checks that $I\overline{\phi_2}\in\Vv$ is also in this orthogonal complement, and by the same argument as above linearly independent of $\phi_2$. Iterating this procedure one obtains an even dimensional basis of $\Vv$.
\hfill $\Box$

\vspace{.2cm}

\noindent {\bf Proof} of the Theorem~\ref{theo-evensym}. Let us begin by diagonalizing $T^*T=U^*MU$. The set $N=\overline{U}JTU^*$. As above one checks $N$ is normal, but now rather symmetric than skew-symmetric. Then let us decompose $N=N_1+\imath N_2$ where $N_1=\frac{1}{2}(N+\overline{N})$ and $N_2=\frac{1}{2\imath}(N-\overline{N})$. Similar as in the proof of Proposition~\ref{prop-normalskew}, $N_1$ and $N_2$ are commuting self-adjoints which are now real. Thus there exists an orthogonal operator $O$ diagonalizing both of them:
$$
O\,N_1\,O^t\;=\;M_1\;,
\qquad
O\,N_2\,O^t\;=\;M_2\;,
$$
where $M_1$ and $M_2$ are real multiplication operators in the spectral representation. Thus 
$$
T\;=\;JU^tNU
\;=\;
JU^tO^t(M_1+\imath\,M_2)OU
\;=\;
JA^tJA
\;,
$$
where $A=O'(M_1+\imath\,M_2)^{\frac{1}{2}}OU$ with $O'$ as in Proposition~\ref{prop-standardforms}.

\vspace{.2cm}

Next let us show that for $T\in\FM(\Hh,J)$ there exists a finite dimensional partial isometry $V\in\BM(\Hh,J)$ such that $T+V$ is invertible. Indeed $\Ker(T^*)=J\Cc\Ker(T)$, so if $(\phi_n)_{n=1,\ldots,N}$ is an orthonormal basis of $\Ker(T)$, then $(J\,\overline{\phi}_n)_{n=1,\ldots,N}$ is an orthonormal basis of $\Ker(T^*)$. Let us set $V=\sum_{n=1}^N J\,|\overline{\phi}_n\rangle\langle\phi_n|$. From this point on, all the arguments are very similar to those in the proof of Theorem~\ref{theo-Z2}.
\hfill $\Box$

\section{Odd symmetric Noether-Gohberg-Krein theorem}
\label{sec-oddGK}

The object of this section is to given an example of a index theorem connecting the $\ZM_2$-index of an odd symmetric Fredholm operator to a topological $\ZM_2$-invariant, simply by implementing an adequate symmetry in the classical Noether-Gohberg-Krein theorem. Let $\Hh$ be a separable complex Hilbert space with a real unitary $I$ satisfying $I^2=-\one$. The set of unitary operators on $\Hh$ having essential spectrum $\{1\}$ is denoted by $\UM_\ess(\Hh)$. Further let $\SM^1=\{z\in\CM\,|\,|z|=1\}$ denote the unit circle. Focus will be on continuous function $f\in C(\SM^1,\UM_\ess(\Hh))$ for which the eigenvalues are continuous functions of $z\in\SM^1$ by standard perturbation theory. Each such function $f\in C(\SM^1,\UM_\ess(\Hh))$ has a well-defined integer winding number which can be calculated as the spectral flow of the eigenvalues of $t\in[0,2\pi)\mapsto f(e^{\imath t})$ through $-1$ (or any phase $e^{\imath\varphi}$ other than $1$), counting passages in the positive sense as $+1$, and in the negative sense as $-1$. It is well-known ({\it e.g.} \cite{Phi}) that the winding number labels the connected components of $C(\SM^1,\UM_\ess(\Hh))$ and establishes an isomorphism between the fundamental group of $\UM_\ess(\Hh)$ and $\ZM$. Furthermore, the Noether-Gohberg-Krein theorem \cite{Noe,GK,BS} states that the winding number is connected to the Fredholm index of the Toeplitz operator associated to $f$. The construction of the Toeplitz operator is recalled below. A precursor of this theorem was proved by F. Noether in the first paper exhibiting a non-trivial index \cite{Noe}. Before going on, let us point out that instead of $\UM_\ess(\Hh)$ as defined above, one can also work with the set of invertibles on $\Hh$ for which there is path from $0$ to $\infty$ in the complement of the essential spectrum (defined as the complement of the discrete spectrum). Indeed, using Riesz projections these cases reduce to the above and the spectral flow is calculated by counting the passages by the above path. Let us point out that also this set of invertibles is compactly stable as can be shown using analytic Fredholm theory.

\vspace{.2cm}

Now an odd symmetry will be imposed on the function $f$, namely
\begin{equation}
\label{eq-GKsym}
I^*\,f(\overline{z})\,I
\;=\;
f(z)^t
\;=\;
\overline{f(z)^{-1}}
\;.
\end{equation}
where in the second equality the unitarity of $f(z)$ was used. As the real points $z=1$ and $z=-1$ are invariant under complex conjugation, \eqref{eq-GKsym} implies a condition for the unitaries $f(1)$ and $f(-1)$, namely they are odd symmetric (if $\Hh$ is finite dimensional, this means that they are in Dyson's symplectic circular ensemble). Such an odd symmetric unitary operator $u$  has a Kramer's degeneracy so that each eigenvalue has even multiplicity (this follows from Proposition~\ref{prop-finitecase} , but is well-known for unitary operators).   Furthermore, by \eqref{eq-GKsym} the spectra of $f(z)$ and $f(\overline{z})$ are equal. Schematic graphs of the spectra of $t\in[-\pi,\pi]\mapsto f(e^{\imath t})$ are plotted in Figure~1. One conclusion is that the winding number of $f$ vanishes (of course, this follows by a variety of other arguments). On the other hand, contemplating a bit on the graphs one realizes that there are two distinct types of graphs which cannot be deformed into each other: the set of spectral curves with Kramers degeneracy at $t=0$ and $t=\pi$ and reflection symmetry at $t=0$ has two connected components. Let us denote by $\mbox{\rm Wind}_2(f)\in\ZM_2$ the homotopy invariant distinguishing the two components, with $0$ being associated to the trivial component containing $f=\one$. One way to calculate $\mbox{\rm Wind}_2(f)$ is to choose $\varphi\in (0,2\pi)$ such that $e^{\imath\varphi}$ is not in the spectrum of $f(1)$ and $f(e^{\imath \pi})$; then the spectral flow of $t\in[0,\pi)\mapsto f(e^{\imath t})$ by $e^{\imath \varphi}$ modulo $2$ (or simply the number of crossings by $e^{\imath \varphi}$ modulo $2$) is $\mbox{\rm Wind}_2(f)$. This allows to read off $\mbox{\rm Wind}_2(f)$ for the examples in Figure~1.

\begin{figure}
\begin{center}
\includegraphics[height=5.7cm]{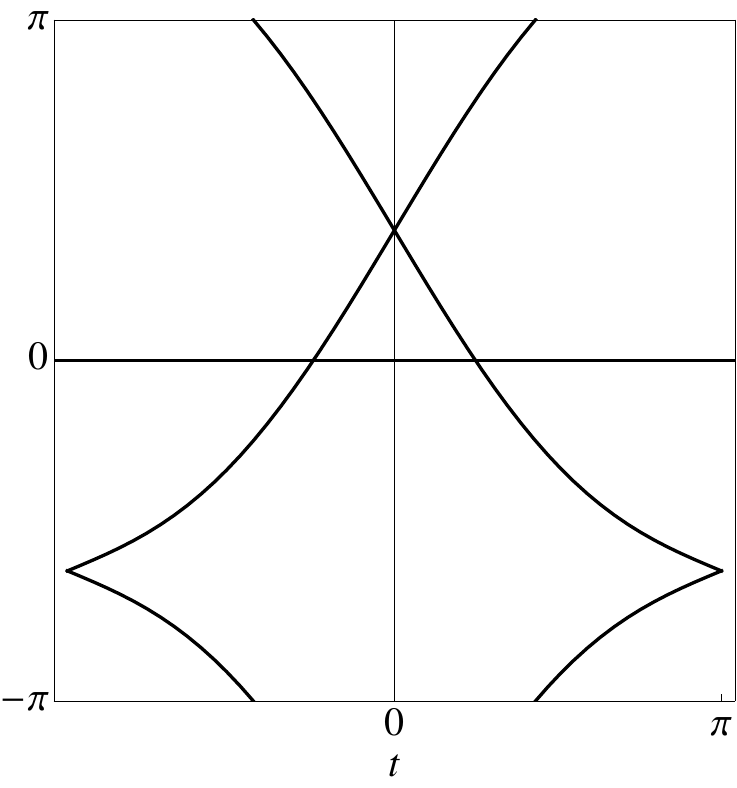}
\hspace{-.1cm}
\includegraphics[height=5.7cm]{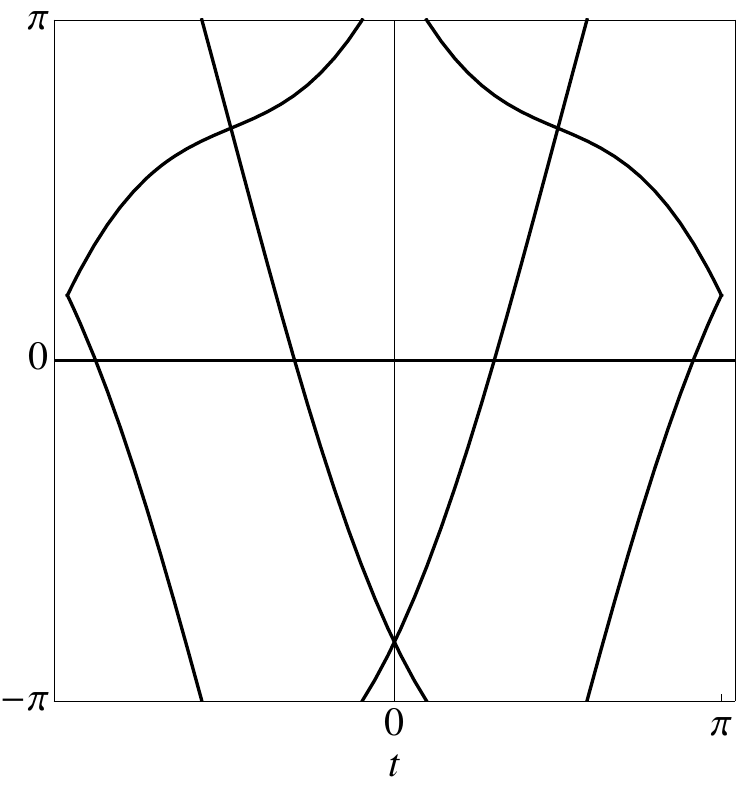}
\hspace{-.1cm}
\includegraphics[height=5.7cm]{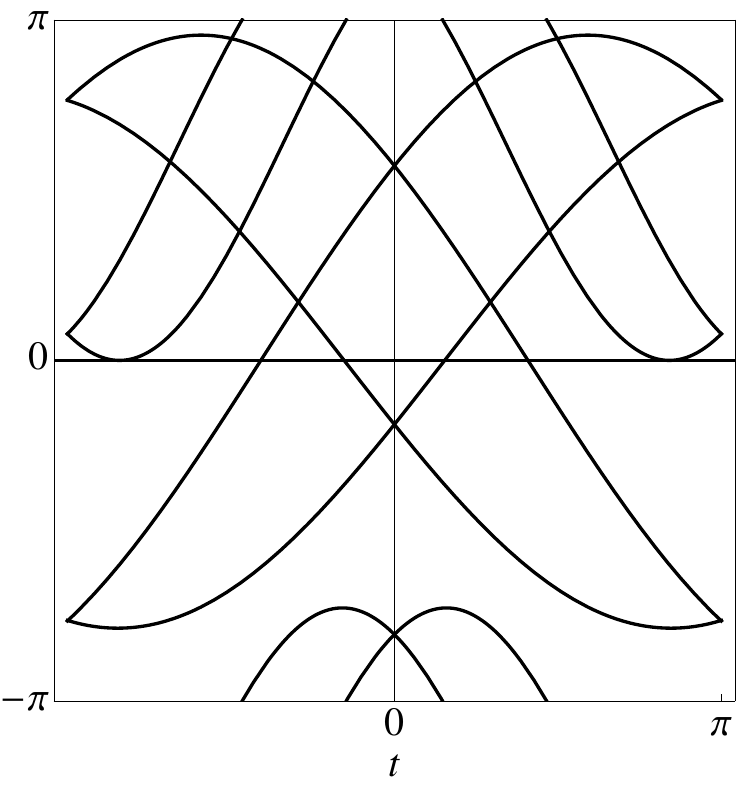}
\caption{\it Schematic representation of the phases of the eigenvalues of $t\in[-\pi,\pi]\mapsto f(e^{\imath t})$ for three examples with the symmetry {\rm \eqref{eq-GKsym}}. The first one is non-trivial, that is $\mbox{\rm Wind}_2(f)=1$, and can actually be seen to be a perturbation of the Fourier transform of $S\oplus S^*$, while the other two both have  $\mbox{\rm Wind}_2(f)=0$. The reader is invited to find the corresponding homotopy to a constant $f$ in the latter two cases.}
\end{center}
\end{figure}

\vspace{.2cm}

The aim is to calculate $\mbox{\rm Wind}_2(f)$ as the $\ZM_2$-index of the Toeplitz operator $T_f$ associated to $f$. The operator $T_f$ turns out to be odd symmetric w.r.t. an  adequate real skew-adjoint unitary. Let us recall the construction of $T_f$. First one considers $f$ as an operator on the Hilbert space $L^2(\SM^1)\otimes\Hh$ where $L^2(\SM^1)$ is defined using the Lebesgue measure on $\SM^1$:
$$
(f\psi)(z)\;=\;f(z)\psi(z)
\;.
$$
On $L^2(\SM^1)$ one has the Hardy projection $P$ onto the Hardy space $H^2$ of positive frequencies. Its extension $P\otimes\one$ to $L^2(\SM^1)\otimes\Hh$ is still denoted by $P$. The discrete Fourier transform $\Ff:L^2(\SM^1) \to\ell^2(\ZM)$ is an Hilbert space isomorphism, under which $f$ and $P$ become operators on $\ell^2(\ZM)\otimes \Hh$ that will be denoted by the same letters. In this representation, $P$ is the projection onto the subspace $\ell^2(\NM)\subset\ell^2(\ZM)$ which is isomorphic to $H^2$. Now the Toeplitz operator on $\Hh'=\ell^2(\NM)\otimes\Hh$ is by definition
$$
T_f\;=\;PfP
\;.
$$
This is known \cite{BS} to be a Fredholm operator (for continuous $f$) and its index is equal to (minus) the winding number of $f$. On the Hilbert space $L^2(\SM^1)\otimes\Hh$ a real skew-adjoint unitary is now defined by
$$
(I'\psi)(z)\;=\;I\,\psi(\overline{z})
\;,
\qquad
\psi\in\Hh'
\;.
$$
As it commutes with $P$, this also defines real skew-adjoint unitary $I'$ on $\Hh'=\ell^2(\NM)\otimes\Hh$. It is a matter of calculation to check that the odd symmetry \eqref{eq-GKsym} of $f$ is equivalent to
$$
(I')^*\,(T_f)^t\,I'\;=\;T_f
\;.
$$
Thus Theorem~\ref{theo-Z2} applied to $\Hh'$ furnished with $I'$ assures the existence of $\Ind_2(T_f)$.

\begin{theo}
\label{theo-Z2GK} 
One has $\mbox{\rm Wind}_2(f)=\Ind_2(T_f)$ for all $f\in C(\SM^1,\UM_\ess(\Hh))$.
\end{theo}

Let us give some non-trivial examples. Let $\Hh=\CM^2$. For $n\in\ZM$, consider the function
$$
f_n(z)\;=\;
\begin{pmatrix}
z^n & 0 \\ 0 & \overline{z}^n
\end{pmatrix}
$$
written in the grading of $I=\left(\begin{smallmatrix} 0 & -\one \\ \one & \;\;0 \end{smallmatrix}\right)$. Then $f_n$ satisfies \eqref{eq-GKsym}. The associated Toeplitz operator on $\Hh=\ell^2(\NM)\otimes\CM^2$ is
$$
T_{f_n}\;=\;
\begin{pmatrix}
S^n & 0 \\ 0 & (S^*)^n
\end{pmatrix}
\;,
$$
where $S:\ell^2(\NM)\to\ell^2(\NM)$ is the left shift. One readily checks separately that indeed $\mbox{\rm Wind}_2(f)=n\,\mbox{mod}\,2$ and $\Ind_2(T_f)=n\,\mbox{mod}\,2$. Now Theorem~\ref{theo-Z2GK} follows for the case $\Hh=\CM^2$ from the homotopy invariance of both quantities appearing in the equality, and the general case follows by approximation arguments. It is a fun exercise to write out the explicit homotopy from $T_{f_2}$ to the identity, by following the proof of Theorem~\ref{theo-Z2}(i).

\section{Time-reversal symmetric topological insulators}
\label{sec-topins}

The aim of this short section is to indicate how the $\ZM_2$-index can be used to distinguish different phases of quantum mechanical systems of independent particles described by a bounded one-particle Hamiltonian $H=H^*$ acting on the Hilbert space $\Hh=\ell^2(\ZM^2)\otimes\CM^N\otimes\CM^{2s+1}$. Here $\ZM^2$ models the physical space by means of a lattice (in the so-called tight-binding representation), $\CM^N$ describes internal degrees of freedom over every lattice site except for the spin $s\in\frac{1}{2}\NM$ which is described by $\CM^{2s+1}$. On the spin fiber $\CM^{2s+1}$ act the spin operators $s^x$, $s^y$ and $s^z$ which form an irreducible representation of dimension $2s+1$ of the Lie algebra su$(2)$. It is supposed to be chosen such that $s^y$ is real. Then the time-reversal operator on $\Hh$ is given by complex conjugation followed by a rotation in spin space by $180$ degrees:
$$
I_{\mbox{\rm\tiny s}}
\;=\;
\one\otimes e^{\imath \pi s^y}
\;.
$$
This operator satisfies $I_{\mbox{\rm\tiny s}}^2=-\one$ if $s$ is half-integer, and $I_{\mbox{\rm\tiny s}}^2=\one$ if $s$ is integer. In both cases, the time-reversal symmetry of the Hamiltonian then reads
$$
I_{\mbox{\rm\tiny s}}^*\,\overline{H}\,I_{\mbox{\rm\tiny s}}
\;=\;
H
\qquad
\Longleftrightarrow
\qquad
I_{\mbox{\rm\tiny s}}^*\,H^t\,I_{\mbox{\rm\tiny s}}
\;=\;
H\;,
$$
namely the Hamiltonian is an odd or even symmetric operator pending on whether the spin $s$ is half-integer or integer. This implies that any real function $g$ of the Hamiltonian also satisfies $I_{\mbox{\rm\tiny s}}^*g(H)^tI_{\mbox{\rm\tiny s}}=g(H)$. Here the focus will be on Fermions so that it is natural to consider the Fermi projection $P=\chi(H\leq E_F)$ corresponding to some Fermi energy $E_F$. These Fermions can have an even or odd spin (this is not a contradiction to fundamental principles because the spin degree of freedom can, for example, be effectively frozen out by a strong magnetic field). Then $P$ is either odd or even symmetric. 

\vspace{.2cm}

Up to now, the spatial structure played no role. Now, it is supposed that $H$ is short range in the sense that it has non-vanishing matrix elements only between lattice sites that are closer than some uniform bound. Further let $X_1$ and $X_2$ be the two components of the position operator on $\ell^2(\ZM^2)$, naturally extended to $\Hh$. Then let us consider the operator 
$$
T_P\;=\;PFP\,+\,(\one-P)
\;,
\qquad
F\;=\;\frac{X_1+\imath\, X_2}{|X_1+\imath\, X_2|}
\;,
$$
which is then also odd or even symmetric. The operator $F$ is called the Dirac phase and it is associated to an adequate even Fredholm module. It can be shown \cite{BES} that $PFP$ is a Fredholm operator on $P\Hh$ provided that the matrix elements of $P$ decay sufficiently fast in the eigenbasis of the position operator (more precisely, $|\langle n|P|m\rangle|\leq C(1+|n+m|)^{-(2+\epsilon)}$ is needed). This holds if $E_F$ lies in a gap of the spectrum of $H$, but also if $E_F$ lies in a spectral interval of so-called dynamical Anderson localization \cite{BES}. As $T_P$ is the direct sum of the operators $PFP$ and $\one-P$ on the Hilbert spaces $P\Hh$ and $(\one-P)\Hh$ respectively and $\one-P$ is simply the identity on the second fiber, it follows that $T_P$ is also Fredholm and has the same Noether index as $PFP$. This index is then equal to the Chern number of $P$ which is of crucial importance for labeling the different phases of the integer quantum Hall effect \cite{BES}. Moreover, if $H=(H_\omega)_{\omega\in\Omega}$ is a covariant family of Hamiltonians (namely, $\Omega$ is a compact topological space equipped  with a $\ZM^2$ action such that, for a given projective unitary representation $a\in\ZM^2\mapsto U_a$ of $\ZM^2$, one has $U_aH_\omega U_a^*=H_{a\cdot \omega}$, see \cite{BES} for details), then the index of $T_P$ is almost surely constant w.r.t. to any invariant and ergodic probability measure $\PM$ on $\Omega$.

\vspace{.2cm}

Here the focus will rather be on a time-reversal symmetric Hamiltonian for which thus the Noether index of $T_P$ vanishes. Such Hamiltonians describe certain classes of so-called topological insulators and the prime example falling in the framework described above is the Kane-Mele Hamiltonian \cite{KM} which is analyzed in great detail in \cite{ASV}. It has odd time-reversal symmetry and the associated Fermi projection (for a periodic model and $E_F$ in the central gap) was shown to be topologically non-trivial for adequate ranges of the parameters \cite{KM,ASV}. While here the model dependent calculation of the associated $\ZM_2$-index is not carried out, the following result is nevertheless in line with these findings. It also shows that the $\ZM_2$-index can be used to distinguish different phases and that the localization length has to diverge at phase transitions, in agreement with the numerical results of \cite{Pro2}.

\vspace{.1cm}

\begin{theo}
\label{theo-topins} 
Consider the Fermi projection $P=(P_\omega)_{\omega\in\Omega}$ of a covariant family of time-reversal invariant Hamiltonians $H=(H_\omega)_{\omega\in\Omega}$ corresponding to a Fermi energy $E_F$ lying in a region of dynamical Anderson localization. Set $T_{P,\omega}=P_\omega FP_\omega+(\one-P_\omega)$. If the spin is half-integer, then the $\ZM_2$-index $\Ind_2(T_{P,\omega})$ is well-defined, $\PM$-almost surely constant in $\omega$  and a homotopy invariant w.r.t. norm continuous changes of the Hamiltonian respecting the time-reversal symmetry and changes of the Fermi energy, as long as the Fermi energy remains in a region of Anderson localization. 
\end{theo}

\noindent {\bf Proof.} By \cite{BES},  $T_{P,\omega}$ is almost surely a Fredholm operator, which by the above has a vanishing Noether index, but also a well-defined $\ZM_2$-index. Moreover, the difference $T_{P,a\cdot\omega}-U_aT_{P,\omega}U_a^*$ is a compact operator (because $U_aFU_a^*-F$ is compact, note also that $U_a$ is not projective due to the absence of magnetic fields). Hence by Theorem~\ref{theo-Z2}(ii) and the unitary invariance of $\Ind_2$, $\Ind_2(T_{P,\omega})$ is constant along orbits and thus $\PM$-almost surely constant by ergodicity. Now remains to show the homotopy invariance. We suppress the index $\omega$ in the below and follow \cite{Pro2} by noting $\Ind_2(T_P)=\Ind_2(T'_P)$ where $T'_P=g(H)Fg(H)+\hat{g}(H)^2$ is obtained using smooth non-negative functions $g$ and $\hat{g}$ with $\mbox{\rm supp}(g)=(-\infty,E_F]$ and $\mbox{\rm supp}(\hat{g})=[E_F,\infty)$. Indeed, then $g(E_F)=0=\hat{g}(E_F)$.  Furthermore, $E_F$ is $\PM$-almost surely not an eigenvalue of the  Hamiltonian, due to Anderson localization. Therefore one has $g(H)P=g(H)$ and $\hat{g}(H)(\one-P) =\hat{g}(H)$ almost surely, and $G(H)=g(H)+\hat{g}(H)$ has almost surely a trivial kernel and its range is all of $\Hh$. As $T'_P=G(H)T_PG(H)$ the equality of the almost sure $\ZM_2$-indices follows. As $T'_P$ is constructed using smooth functions of the Hamiltonian, it is now possible to make norm continuous deformations of the Hamiltonian and then appeal to Theorem~\ref{theo-Z2}(iii) to conclude the proof.
\hfill $\Box$

\vspace{.2cm}

If the spin is integer, then the operators $T_P$ can be homotopically deformed to the identity (within the class of time-reversal symmetric operators). This is in line with the belief that there are no non-trivial topological insulator phases for two-dimensional Hamiltonians with even time-reversal symmetry.

\vspace{.2cm}

In the remainder of the paper, the implications of a non-trivial $\ZM_2$-invariant for odd time-reversal symmetric systems is discussed. In fact, it seems to be unknown whether $\Ind_2(T_P)$ can be directly measured, but it is believed \cite{KM} that $\Ind_2(T_P)=1$ implies the existence of edge modes that are not susceptible to Anderson localization. Indeed, dissipationless edge transport was shown to be robust under the assumption of non-trivial spin Chern numbers \cite{Sch}. Theorem~\ref{theo-topins2} below shows that this assumption holds if $\Ind_2(T_P)=1$.

\vspace{.2cm}

Spin Chern numbers for disordered systems were first defined by Prodan \cite{Pro}. Let us review their construction in a slightly more general manner that is possibly applicable to other models. Suppose given another bounded self-adjoint observable $A=A^*\in\BM(\Hh)$ which is odd skew-symmetric, namely $I^*A^tI=-A$. Associated with $A$ and the Fermi projection $P$ is the self-adjoint operator $PAP$ which is also odd skew-symmetric. The spectrum of both $A$ and $PAP$ is odd, that is $\sigma(PAP)=-\,\sigma(PAP)$. It will now be assumed that $0$ is not in the spectrum of $PAP$ when viewed as operator on $P\Hh$. This allows to define two associated Riesz projections $P_\pm$ by taking contours $\Gamma_\pm$ around the positive and negative spectrum of $PAP$:
%
$$
P_\pm
\;=\;
\oint_{\Gamma_\pm}\frac{dz}{2\pi\imath}\;(z-PAP)^{-1}
\;.
$$
One then has $P=P_++P_-$ and $P_+P_-=0$ and, most importantly, $I^*(P_\pm)^tI=P_\mp$. Therefore $P_\pm$ provide a splitting of $P\Hh$ into two subspaces $P_+\Hh$ and $P_-\Hh$  which are mapped onto each other under the time-reversal operator $I\Cc$. If now the matrix elements of $P$ in the eigenbasis $|n\rangle$ of the position operator has decay as described above and also $A$ has such decay ({\it e.g.}, $A$ is a local operator), then one can show that also the matrix elements of $P_\pm$ decay ({\it e.g.} by following the arguments in \cite{Pro} imitating those leading to the Combes-Thomas estimate). Consequently $[F,P_\pm]$ is compact and therefore $P_\pm FP_\pm$ are Fredholm operators on $P_\pm \Hh$ with well-defined Noether indices which, by the arguments in the proof of Theorem~\ref{theo-topins2}  below, satisfy $\Ind(P_+ FP_+)=-\,\Ind(P_- FP_-)$. Under adequate decay assumptions these indices are again equal to the Chern numbers of $P_\pm$. What is now remarkable is that the indices $\Ind(P_\pm  FP_\pm)$ are also stable under perturbations which break time-reversal invariance, such as magnetic fields. Hence Theorem~\ref{theo-topins2} below shows that a non-trivial $\ZM_2$-invariant defined for a time-reversal invariant system leads, under adequate hypothesis, to non-trivial invariants that are stable also if time-reversal symmetry  is broken.

\vspace{.2cm}

All the above hypothesis on $A$ hold for the Kane-Mele model with small Rashba coupling if $A=s^z$ is the $z$-component of the spin operator. In this situation the Chern numbers of $P_\pm$ are then called the spin Chern numbers \cite{Pro,ASV,Sch}.

\begin{theo}
\label{theo-topins2} 
Consider the Fermi projection of a time-reversal invariant Hamiltonian $H$ corresponding to a Fermi energy $E_F$ lying in a region of dynamical Anderson localization. Suppose that $A$ is a self-adjoint operator such that $0$ is not in the spectrum of $PAP\in\BM(P\Hh)$ and that for the Riesz projections $P_+$ and $P_-$ on the positive and negative spectrum of $PAP$, the commutators $[F,P_\pm]$ are compact. Then $\Ind_2(T_P)=\Ind(P_\pm FP_\pm)\,\mbox{\rm mod}\,2$.
\end{theo}

\noindent {\bf Proof.} Let $\Ind(P_+ FP_+)=k$. We show that there is a homotopy within $\FM(\Hh,I)$ connecting the operator $T_P$ to an operator $T_0$ with $\Ind_2(T_0)=k\,\mbox{\rm mod}\,2$. Let us begin by choosing an orthonormal basis $(\phi_n)_{n\in\NM}$ in the Hilbert space $P_+\Hh$. Then $\Phi=(\phi_1,\phi_2,\ldots):\ell^2(\NM)\to P_+\Hh$ is a Hilbert space isomorphism. Let the standard complex conjugation on $\ell^2(\NM)$ also be denoted by $\Cc$, and set $\overline{\Phi}=\Cc\Phi\Cc$. Because $I^*(P_+)^tI=P_-$, also $I\overline{\Phi}:\ell^2(\NM)\to P_-\Hh$ is a Hilbert space isomorphism and so is $(\Phi,I\overline{\Phi}):\ell^2(\NM)\oplus \ell^2(\NM)\to P\Hh$. Also consider $I=\left(\begin{smallmatrix} 0 & -\one \\ \one & \;\;0 \end{smallmatrix}\right)$ as an operator on $\ell^2(\NM)\oplus \ell^2(\NM)$.  Then $I\Cc(\Phi,I\overline{\Phi})=(\Phi,I\overline{\Phi})I\Cc$. As above, let the left shift on $\ell^2(\NM)$ be denoted by $S$. Then $G_0=\Phi S^k\Phi^*$ is a Fredholm operator on $P_+\Hh$ with $\Ind(G_0)=k$. Hence there exists a homotopy $s\in[0,1]\mapsto G_s\in\FM(P_+\Hh)$ from  $G_0$ to $G_1=P_+ FP_+$. Extending $G_s$ by $0$ to all $\Hh$, we next define $T_s=G_s+I^*(G_s)^tI+(\one-P)$. By construction, $T_s\in\FM(\Hh,I)$. Furthermore, $T_0=(\Phi,I\overline{\Phi})\left(\begin{smallmatrix} S^k & 0\\ 0 & (S^*)^k \end{smallmatrix}\right)(\Phi,I\overline{\Phi})^*+\one-P$, so that indeed $\Ind_2(T_0)=k\,\mbox{\rm mod}\,2$, see Section~\ref{sec-oddGK}. On the other hand, $T_1=P_+ FP_++P_- FP_-+(\one-P)$. Next $P_\pm FP_\mp=P_\pm[F,P_\mp]$ is compact and odd symmetric. It follows that also $s\in[1,2]\mapsto T_1+(s-1)(P_+FP_-+P_-FP_+)$ is a homotopy in $\FM(\Hh,I)$. As $T_2=T_P$, the proof is completed.
\hfill $\Box$

\vspace{.3cm}

\noindent {\bf Acknowledgements.} The author thanks Maxim Drabkin and Giuseppe De Nittis  for comments and proof reading. This work was partially funded by the DFG. 





\end{document}